\documentclass[preprint,aps,nofootinbib]{revtex4-1}
\pdfoutput=1

\usepackage{amsmath}  
\usepackage{amsbsy,amssymb,amsfonts}
\usepackage{graphicx}
\usepackage{color}
\usepackage{epsf}
\usepackage[utf8x]{inputenc}
\usepackage{hyperref}
\usepackage{multirow}
\def\bravert{\egroup\,\vrule\,\bgroup}

{\catcode`\|=\active
  \gdef\Twoint#1{\left(\mathcode`\|"8000\let|\bravert {#1}\right)}}
{\catcode`\|=\active
  \gdef\Braket#1{\left<\mathcode`\|"8000\let|\bravert {#1}\right>}}

\newcommand{\beq}{\begin{equation}}
\newcommand{\eeq}{\end{equation}}
\newcommand{\beqa}{\begin{eqnarray}}
\newcommand{\eeqa}{\end{eqnarray}}
\newcommand{\bea}{\begin{array}}
\newcommand{\eea}{\end{array}}

\newcommand{\bef}{\begin{figure}}
\newcommand{\ef}{\end{figure}}
\newcommand{\bc}{\begin{center}}
\newcommand{\ec}{\end{center}}
\newcommand{\bt}{\begin{table}}
\newcommand{\et}{\end{table}}
\newcommand{\btb}{\begin{tabular}}
\newcommand{\etb}{\end{tabular}}

\newcommand{\mea}{{\emph{et al.}}}

\newcommand{\au}{{\em a.u.}}

\begin{document}

\title {${\cal{(P,T)}}$-Odd Tensor-Pseudotensor Interactions \\
         in atomic {$^{199}$Hg} and {$^{225}$Ra} }

\vspace*{1cm}

\author{Timo Fleig}
\email{timo.fleig@irsamc.ups-tlse.fr}
\affiliation{Laboratoire de Chimie et Physique Quantiques,
             IRSAMC, Universit{\'e} Paul Sabatier Toulouse III,
             118 Route de Narbonne, 
             F-31062 Toulouse, France }
\date{\today}

\vspace*{1cm}
\begin{abstract}
Highly correlated pure {\it{ab initio}} relativistic configuration interaction theory is in the
present paper applied to the calculation of the tensor-pseudotensor ${\cal{(P,T)}}$-violating
nucleon-electron interaction constant in the electronic ground states of atomic mercury and radium. 
The final best obtained results are $R_T$(Hg) = $-4.43$\, [$10^{-20}$ $\left<\sigma_N\right>$ $e$ cm]
and $R_T$(Ra) = $-15.0$\, [$10^{-20}$ $\left<\sigma_N\right>$ $e$ cm]. The accuracy of the employed
electronic-structure models are confirmed by determining the static electric dipole polarizability
$\alpha_d$(Hg) = $35.7$ {\it{a.u.}} which is in accord with the experimental value to about
$5$\%. $R_T$(Ra) will be useful for constraining (or obtaining) the ${\cal{(CP)}}$-violating parameter 
$C_T$ when combined with future measurements of the electric dipole moment of the radium atom.
\end{abstract}

\maketitle
\section{Introduction}
\label{SEC:INTRO}

Charge-Parity (${\cal{(CP)}}$)-violation has so far been observed in nature only in the decays of
certain heavy mesons -- such as the $K$ and $B$ mesons \cite{K-meson,Abe:2001xe,B-meson} -- which are
flavor-changing processes driven by the weak interaction. These sources of ${\cal{(CP)}}$-violation
have become an integral part of the Standard Model (SM) of particle physics through the 
Cabibbo-Kobayashi-Maskawa (CKM) quark-mixing matrix \cite{Kobayashi}. However, the observed 
disparity of matter and antimatter in the universe \cite{Dine_Kusenko_MatAntimat2004} which requires
${\cal{(CP)}}$ symmetry to be violated \cite{Sakharov_JETP1967} cannot be explained solely through 
SM ${\cal{(CP)}}$-violation. Additional sources of ${\cal{(CP)}}$-violation are predicted by most
theory models Beyond the SM \cite{ramsey-musolf_review2_2013}, and they give rise to electric dipole 
moments (EDM) in atomic matter many orders of magnitude larger than the SM predictions 
\cite{hoogeveen_SMeEDM1990}.

If one accepts that ${\cal{CPT}}$ symmetry \cite{pauli_lorentz_CPT} is unbroken -- and this is strongly 
suggested by its intertwining with Lorentz invariance -- then ${\cal{(CP)}}$-violation implies 
${\cal{T}}$-violation, which means that, e.g., an atomic energy shift would change sign if the 
laboratory were to travel backwards in time. The search for additional ${\cal{(CP)}}$-violation in
nature, therefore, can be carried out as search for EDMs of elementary particles and search for
a ${\cal{T}}$-violating piece of the weak interaction, for instance among the hadrons and leptons of 
an atomic system \cite{Herczeg_PRD2003}. The latter can arise from the tree-level exchange of a BSM 
mediator particle such as in the framework of leptoquark scenarios 
\cite{Fuyuto_Leptoquark_ArXiV2018,Herczeg_PRD2003}.

An important potential manifestation of ${\cal{(CP)}}$-violation in atomic systems with closed electronic
shells is an EDM due to the semi-leptonic spin-dependent nucleon-electron 
tensor-pseudotensor (Ne-TPT) interaction \cite{khriplovich_lamoreaux}. This is due to the zero total 
electronic angular momentum ${\bf{J}}$ in the closed-shell atomic ground state, and therefore the atomic 
EDM depending on nuclear angular momentum only. The Ne-TPT interaction is also present in open-shell 
systems, but there the scalar-pseudoscalar nucleon-electron interaction is by far dominant for heavier 
nuclei due to its scaling with nucleon number. 

Experimental upper bounds to the EDMs of closed-shell atoms may therefore be used to constrain the possible
value of the fundamental Ne-TPT coupling parameter $C_T$. Such experiments have been carried out in the
past in particular on the Hg atom \cite{Heckel_Hg_PRL2016,swallows_Hg_PRA2013} and the Ra atom 
\cite{bishof_Ra_PRC2016,parker_Ra_PRL2015}, among others \cite{sato_Xe_HFINT2015}. These measurements
are subject to continued improvements, and the same should hold true for the electronic-structure
calculations required to interpret them in terms of fundamental ${\cal{(CP)}}$-violating parameters.

In the present paper I focus on the Ne-TPT atomic interaction constants of interest for the EDM 
measurements on {$^{199}$Hg} and {$^{225}$Ra}. The specific aims of the present work are threefold:
First, a relativistic correlated electronic-structure method that has been applied earlier to the
calculation of atomic \cite{FleigJung_JHEP2018} and molecular \cite{PhysRevA.96.040502} EDMs and 
${\cal{P,T}}$-odd enhancements is made available for the calculation of the Ne-TPT interaction
in both atomic and molecular systems. The corresponding theory is outlined in the following section
of the paper, sec. \ref{SEC:THEORY}. Second, the approach is put to the test on {$^{199}$Hg}
where results for the Ne-TPT interaction constant have been obtained earlier using Coupled Cluster
methods \cite{sahoo_CPodd-Hg_PRD2017,singh_CPodd-Hg_PRA2015}. Third, predictions for the Ne-TPT 
interaction constant in {$^{225}$Ra} are made using highly-correlated atomic wavefunctions which
are compared with present theory results for this atom \cite{dzuba_flambaum_PRA2009}.
All applications are discussed in section \ref{SEC:APPL}.
\section{Theory}
\label{SEC:THEORY}
An atomic electric dipole moment (EDM) resulting solely from T-PT nucleon-electron interaction is defined as
\begin{equation}
 \label{EQ:ATEDM}
 d_a = R_T C_T
\end{equation}
with $C_T$ the fundamental Ne-TPT coupling constant and $R_T$ the corresponding atomic interaction 
constant. The latter is determined in the framework of an effective field theory involving a neutral
weak current between electrons and nucleons. This nucleon-electron tensor-pseudotensor interaction 
Hamiltonian has been given as \cite{hinds_loving_sandars1976}
\begin{equation}
 \label{EQ:TPT_HAM}
 \hat{H}_{Ne}^{T-PT} = \frac{\imath G_F}{\sqrt{2}}\, 2\, C_T\, 
                       {\boldsymbol{\gamma}}_e \cdot {\boldsymbol{\sigma}}_N\, \rho({\bf{r}})
\end{equation}
where $G_F$ is the Fermi constant, $\gamma$ is an electronic Dirac matrix,
${\boldsymbol{\sigma}}_N = \left<\sigma_N\right>_{\Psi_N} \frac{\bf{I}}{I}$ with $\Psi_N$ a nuclear spinor,
$\sigma_N$ a Pauli matrix, ${\bf{I}}$ the nuclear spin, and $\rho({\bf{r}})$ is the nuclear density at
position ${\bf{r}}$.

For convenience, the nuclear state is chosen as $\left| I, M_I=I \right>$. From this it follows that
$\left< I, M_I=I \right| \left({{\boldsymbol{\sigma}}_N}\right)_k \left| I, M_I=I \right> = 0$ 
$\forall k \in \{1,2\}$ where $k$ denotes cartesian components. Using the identity
$\left< I, M_I=I \right| \left({{\boldsymbol{\sigma}}_N}\right)_3 \left| I, M_I=I \right> = 
\left<\sigma_N\right>_{\Psi_N}$ in \au\ the above Hamiltonian is rewritten as
\begin{equation}
 \label{EQ:TPT_HAM1}
 \hat{H}_{Ne}^{T-PT} = \frac{\imath G_F}{\sqrt{2}}\, 2\, C_T\, 
 \left(\gamma_e\right)^3\, \left<\sigma_N\right>_{\Psi_N}\, \rho({\bf{r}}).
\end{equation}
where upper indices on four-tensors conventionally correspond to contravariant components.

The determination of the atomic EDM resulting from the corresponding T-PT interaction in
first order in perturbation theory results  calculation of the matrix element
\begin{equation}
 \label{EQ:TPT-ME}
 M_{Ne}^{T-PT} = \left< \psi_e | \imath \left(\gamma_e\right)^3\, \rho({\bf{r}}) | \psi_e \right>
\end{equation}
where $\psi_e$ is the electronic wavefunction of the atomic state under consideration. The implementation
of this matrix element has been carried out in a locally modified version of the \verb+DIRAC+ program 
package \cite{DIRAC16}, using the relation $\left(\gamma_e\right)^k = \gamma^0 \alpha_k = \beta \alpha_k$ 
where $\alpha, \beta$ are Dirac matrices.

The atomic EDM then results from the interaction constant
\begin{equation}
 \label{EQ:RT_DETAIL}
 R_T = \sqrt{2} G_F \left<\sigma_N\right>_{\Psi_N}\, M_{Ne}^{T-PT}
\end{equation}
and Eq. (\ref{EQ:ATEDM}).

\section{T-PT interaction in {$^{225}$Ra} and {$^{199}$Hg}}
\label{SEC:APPL}
\subsection{Technical details}

Atomic basis sets of Gaussian functions are used, denoted valence quadruple-zeta (vQZ) and including all 
available polarizing and valence-correlating functions \cite{dyall_basis_2004,dyall_s-basis}. 
The complete sets amount to (38s,35p,24d,16f,3g) functions for Ra and (34s,30p,19d,13f,4g,2h) functions 
for Hg, respectively. Wavefunctions for the {$^1S_0$} electronic ground state of the respective atoms are
obtained through a closed-shell Hartree-Fock (HF) calculation using the Dirac-Coulomb Hamiltonian
including the external electric field
\begin{eqnarray}
 \nonumber
\hat{H} &:=& \hat{H}^{\text{Dirac-Coulomb}} + \hat{H}^{\text{Int-Dipole}} \\
&=& \sum\limits^N_j\, \left[ c\, \boldsymbol{\alpha}_j \cdot {\bf{p}}_j + \beta_j c^2 
 + \frac{Z}{r_j}{1\!\!1}_4 \right]
     + \sum\limits^N_{j,k>j}\, \frac{1}{r_{jk}}{1\!\!1}_4
        + \sum\limits_j\, {\bf{r}}_j \cdot {\bf{E}_{\text{ext}}}\, {1\!\!1}_4\,,
 \label{EQ:HAMILTONIAN}
\end{eqnarray}
where the indices $j,k$ run over $N$ electrons, $Z$ is the proton number ($N=Z$ for neutral atoms).
Atomic units (\au) are used throughout ($e = m_0 = \hbar = 1$). This approach corresponds to what
other authors \cite{dzuba_flambaum_PRA2009} call Random-Phase Approximation (RPA).
The DCHF calculation is followed by linear expansion in the basis of Slater determinants formed by the 
occupied and virtual set of 4-spinors and diagonalization of the DC Hamiltonian including the external 
electric field in that basis (Configuration Interaction (CI) approach) \cite{knecht_luciparII}. The 
resulting ``correlated'' wavefunctions $\psi_e$ -- where the CI expansion coefficients are fully relaxed 
with respect to the external field -- are then introduced into Eq. (\ref{EQ:TPT-ME}),
and the resulting ${\cal{P}}$-odd expectation value gives the T-PT interaction constant {\it{via}}
Eq. (\ref{EQ:RT_DETAIL}).

The nomenclature for CI models is defined as: S, D, T, etc. denotes Singles, Doubles, Triples etc.
replacements with respect to the closed-shell DCHF determinant. The following number is the number of
correlated electrons and encodes which occupied shells are included in the CI expansion. 
In the case of Hg we have 
$12 \mathrel{\widehat{=}} (5d,6s)$, 
$18 \mathrel{\widehat{=}} (5p,5d,6s)$, 
$32 \mathrel{\widehat{=}} (4f,5p,5d,6s)$, 
$34 \mathrel{\widehat{=}} (5s,4f,5p,5d,6s)$,
$42 \mathrel{\widehat{=}} (4s,4p,5s,4f,5p,5d,6s)$, and 
$44 \mathrel{\widehat{=}} (4d,5s,4f,5p,5d,6s)$.
In the case of Ra these are 
$10 \mathrel{\widehat{=}} (6s,6p,7s)$, 
$20 \mathrel{\widehat{=}} (5d,6s,6p,7s)$,
$28 \mathrel{\widehat{=}} (5s,5p,5d,6s,6p,7s)$,
$36 \mathrel{\widehat{=}} (4s,4p,5s,5p,5d,6s,6p,7s)$, and
$42 \mathrel{\widehat{=}} (4f,5s,5p,5d,6s,6p,7s)$.
The notation type S8\_SD36, as an example, means that the model SD36 has been approximated by
omitting Double excitations from the $(4s,4p)$ shell.

The nuclear spin quantum number is $I=1/2$ both for {$^{225}$Ra} and {$^{199}$Hg}, respectively
\cite{stone_INDC2015}. Atomic nuclei are described by Gaussian charge distributions 
\cite{Visscher_Dyall_nuclcha} with exponents $\zeta_{\text{Hg}} = 1.4011788914 \times 10^8$ and 
$\zeta_{\text{Ra}} = 1.3101367628 \times 10^8$, respectively.

Atomic static dipole polarizabilites are obtained from fitting the total electronic energies using
seven field points of strengths $E_{\rm{ext}} \in \{-1.2, -0.6, -0.3, 0.0, 0.3, 0.6, 1.2\} \times 10^{-4}$
\au\ For calculation of the T-PT interaction constant $E_{\rm{ext}} = 0.3 \times 10^{-4}$ \au\ and 
$E_{\rm{ext}} = 0.6 \times 10^{-4}$ \au\ for Hg and Ra, respectively.

\subsection{Results and discussion}


\begin{table}[h]
 \caption{Ne-TPT interaction constant and static electric dipole polarizability for the Hg atom 
          using different wavefunction models
          \label{TAB:RT_HG} }

 \vspace*{0.5cm}
 \begin{tabular}{l|cc|cc}
 Model/virtual cutoff [\au] & 
 \multicolumn{2}{c|}{$\alpha_d$ [\au]} &
 \multicolumn{2}{c}{$R_T$ [$10^{-20}$ $\left<\sigma_N\right>$ $e$ cm]} \\ \hline\hline
                        &  \multicolumn{2}{c|}{Basis set}  & \multicolumn{2}{c}{Basis set}     \\
                        & vDZ        & vTZ        & vDZ            & vTZ              \\ \hline
        RPA/-           & $44.5$     & $45.5$     & $-4.70$        & $-4.94$          \\ 
        SD12/22         & $35.0$     & $33.5$     & $-4.05$        & $-4.25$          \\
	SD18/22         & $30.7$     & $34.2$     & $-5.52$        & $-5.82$          \\
	SD12\_SDT18/22  &            &            & $-5.35$        &                  \\
	SD32/50         &            &            & $-5.90$        &                  \\
	SD34/22         &            &            & $-4.77$        & $-5.16$          \\
	SD34/50         & $30.0$     & $34.8$     & $-4.95$        & $-5.19$          \\
	S2\_SD34/150    &            &            & $-4.97$        &                  \\
	S8\_SD42/50     &            &            & $-5.14$        &                  \\
	S10\_SD44/50    &            &            & $-5.00$        &                  \\
        SDT12/22        & $37.9$     & $34.4$     & $-3.65$        & $-3.71$          \\
	SDTQ12/22       &            &            & $-3.40$        &                  \\ 
 {\bf{vTZ/SD34/50 +}}$\boldsymbol{\Delta}$ & \multicolumn{2}{c|}{$\boldsymbol{35.7}$} 
                                           & \multicolumn{2}{c}{$\boldsymbol{-4.43}$}  \\ \hline
   Singh \mea\footnote{Ref. \cite{singh_CPodd-Hg_PRA2015}} CCSD$_p$T
   &  \multicolumn{2}{c|}{$34.27$}  & \multicolumn{2}{c}{$-4.30$}       \\
   Dzuba\footnote{Ref. \cite{dzuba_IPpol_superheavy2016}} RPA 
   &  \multicolumn{2}{c|}{$44.9$}  & \multicolumn{2}{c}{ }       \\
   Latha \mea\footnote{Ref. \cite{Latha:2009nq}}   
   &         &          & \multicolumn{2}{c}{$-4.3$}        \\
   Dzuba \mea\footnote{Ref. \cite{dzuba_flambaum_PRA2009}}   
   &         &           & \multicolumn{2}{c}{$-5.1$}        \\ \hline
   Experiment\footnote{Ref. \cite{goebel_hohm_hg1996}} 
   &   \multicolumn{2}{c|}{$33.91$}   &                &
 \end{tabular}
\end{table}


Results from a systematic study on Hg with different wavefunction models are compiled in Table
\ref{TAB:RT_HG}. For the T-PT interaction constant there is a general trend for absolute values, 
independent of basis set: Correlations among the valence electrons (5d and 6s) decreases $R_T$ 
whereas inclusion of inner shells leads to an increase, the sole exception being the 5s shell
which has a strong effect in the opposite direction. The net effect is an increase of about
5\% between RPA and SD34/50 in vTZ basis. Core-valence correlations among the inner (4s,4p,4d)
electrons and valence and subvalence electrons adds another $\approx +4$\% to $R_T$.
To the contrary, replacements higher than Doubles (D) consistently decrease the T-PT interaction 
constant for all investigated shells. Up to Quadruple excitations from the valence (6s,5d) shells
have been considered in the model SDTQ12. The total effect of Triples and Quadruples for the
considered atomic shells is a remarkable $18$\%.

The final value for $R_T$ is obtained from using the SD34/50 result in vTZ basis and by adding 
corrections to this value in the following manner:
\begin{eqnarray*}
\Delta \rightarrow \Delta R_T &=&  R_T({\rm{vDZ/S8\_SD42/50}}) - R_T({\rm{vDZ/SD34/150}}) \\
           & & + R_T({\rm{vDZ/S10\_SD44/50}}) - R_T({\rm{vDZ/SD34/150}}) \\
           & & + R_T({\rm{vTZ/SDT12/22}}) - R_T({\rm{vTZ/SD12/22}}) \\
           & & + R_T({\rm{vDZ/SDTQ12/22}}) - R_T({\rm{vDZ/SDT12/22}}) \\
           & & + R_T({\rm{vDZ/SD12\_SDT18/22}}) - R_T({\rm{vDZ/SD18/22}})
\end{eqnarray*}
$\Delta R_T$ corrects for core-valence correlations from the (4s,4p,4d) shells and for higher
excitations among the valence and the subvalence (5p) electrons.

The final value obtained in this way, vTZ/SD34/50 +$\Delta R_T$, given in Table \ref{TAB:RT_HG}
lines up very well with the results of Singh et al. \cite{singh_CPodd-Hg_PRA2015} (taken without the
small Breit and QED corrections) and Latha et al. \cite{Latha:2009nq}. 
A conservative estimation of the
uncertainty from atomic basis set, higher excitations, and inner-shell correlations yields a
total of $9$\% for $R_T$ by adding estimated individual uncertainties. The true uncertainty is likely
to be significantly smaller because inner-shell effects and higher excitations generally act in 
opposite directions.
This means that the result of Dzuba \mea\ is outside of even the conservative uncertainty bar for the 
present result.
The reason for this could be an incomplete treatment of higher excitations in the CI+MBPT approach
of Dzuba \mea\ since these lead to a large downward correction (on the absolute) of the T-PT
interaction constant for atomic Hg.

For gauging the accuracy of the present electronic structure models the atomic static dipole
polarizability is used for which experimental and other theory results are known. The RPA value
of Dzuba (without correlations) is in excellent agreement with the present RPA value for 
$\alpha_d$ in Hg. However, and as expected, correlation effects are of major importance.
As to be seen in Table \ref{TAB:RT_HG}
for the static electric dipole polarizability of Hg neither basis set effects nor electron correlation
effects behave in a consistent manner, in contrast to those effects on $R_T$. It is, therefore, safer 
to derive the final value for $\alpha_d$ from the largest used basis set only. The correction is in
this case calculated as follows:
\begin{eqnarray*}
\Delta \rightarrow \Delta \alpha_d &=&  \alpha_d({\rm{vTZ/SDT12/22}}) - \alpha_d({\rm{vTZ/SD12/22}})
\end{eqnarray*}
The present final value of $\alpha_d = 35.7$ {\it{a.u.}} is more than $20$\% smaller than the RPA
result and deviates from the experimental value by only about $5$\%. The final value for $R_T$
in Hg has been obtained using a similar computational protocol.


\begin{table}[h]
 \caption{Ne-TPT interaction constant for the Ra atom using different wavefunction models
          \label{TAB:RT_RA} }
 \begin{tabular}{l|ccc}
 Model/virtual cutoff [\au] & \multicolumn{3}{c}{$R_T$ [$10^{-20}$ $\left<\sigma_N\right>$ $e$ cm]} 
        \\ \hline\hline
                                    & \multicolumn{3}{c}{Basis set}     \\
				    & vDZ            & vTZ            & vQZ         \\ \hline
	RPA/-                       & $-14.5$        & $-14.7$       & $-14.7$      \\ 
	SD10/23                     & $-12.5$        & $-13.6$       & $-13.7$      \\ 
	SD10/50                     & $-12.5$        &               &              \\ 
	SD20/23                     & $-13.6$        & $-13.9$       & $-14.0$      \\
	SD28/23                     & $-14.7$        & $-15.0$       & $-14.9$      \\
	SD28/50                     &                & $-14.9$       &              \\
	S8\_SD36/50                 &                & $-15.4$       &              \\
	S14\_SD42/50                &                & $-15.1$       &              \\
	SDT10/23                    & $-11.3$        & $-13.1$       & $-13.2$      \\ 
	SDTQ10/23                   & $-10.7$        &               &              \\ 
 {\bf{vTZ/SD28/50 +}}$\boldsymbol{\Delta R_T}$ & \multicolumn{3}{c}{$\boldsymbol{-15.0}$}         \\ \hline
   Dzuba et al.\footnote{Ref. \cite{dzuba_flambaum_PRA2009}} & \multicolumn{3}{c}{$-18$}  \\ \hline
 \end{tabular}
\end{table}


The general trends for the T-PT interaction constant in Ra, displayed in Table \ref{TAB:RT_RA}, are 
qualitatively the same as in the Hg atom.
Valence correlations and higher excitations than Doubles diminish $R_T$, inner-shell correlations increase
$R_T$, on the absolute. However, in Ra a large basis set effect of nearly $16$\% on higher excitations is 
observed. In the mercury atom the corresponding basis set effect is less than $2$\%, considering only
the change from vDZ to vTZ. 
For this reason the effect of Triples is evaluated with the vQZ basis set for Ra where it is only about
a fourth of that effect with the vDZ basis. Accordingly, the effect of Quadruples will be overestimated
in the vDZ basis, and this correction is, therefore, scaled to the expected value in vQZ basis.

The correction $\Delta R_T$ of the base value for $R_T$ in atomic Ra is obtained from
\begin{eqnarray*}
 \Delta R_T &=&  R_T({\rm{vTZ/S8\_SD36/50}}) - R_T({\rm{vTZ/SD28/50}}) \\
           & & + R_T({\rm{vTZ/S14\_SD42/50}}) - R_T({\rm{vTZ/SD28/50}}) \\
           & & + R_T({\rm{vQZ/SDT10/23}}) - R_T({\rm{vQZ/SD10/23}}) \\
           & & + [R_T({\rm{vDZ/SDTQ10/23}}) - R_T({\rm{vDZ/SDT10/23}})]_{\rm{scaled~to~vQZ}}.
\end{eqnarray*}
The estimated uncertainty of the final result for $R_T$ in atomic Ra is around $8$\%, using the same
approach as for the Hg atom. Again, the result of Dzuba \mea\ is larger than the upper bound to the
present result.

\section{Conclusions}
\label{SEC:CONCL}
The present result for the NeTPT interaction constant can be combined with the recent atomic
EDM measurement of the Argonne group on {$^{225}$Ra} of 
$|d_{{^{225}{\rm{Ra}}}}| < 1.4 \times 10^{-23}\, e$cm \cite{bishof_Ra_PRC2016} to yield a limit on 
the ${\cal{(CP)}}$-violating parameter $C_T$, supposing a single-source interpretation of the atomic 
measurement\footnote{In more sophisticated interpretations such as in Refs.
\cite{Chupp_Ramsey_Global2015,FleigJung_JHEP2018} possible cancellations between leading
${\cal{(CP)}}$-violating contributions to an atomic EDM can be taken into account.}.
From Eq. (\ref{EQ:ATEDM})
\begin{equation}
 |C_T| < \left| \frac{d_{{^{225}{\rm{Ra}}}}}{R_T} \right| 
       = 9.3 \times 10^{-5}\, \frac{1}{\left<\sigma_N\right>}
\end{equation}
The determination of $\left<\sigma_N\right>$ {\it{via}} a simple spherical shell model of the
Ra nucleus, following Ref. \cite{dzuba_flambaum_PRA2009}, is not attempted here due to the known
strong deformation of the {$^{225}$Ra} nucleus. However, it can be assumed that since
$I$({$^{225}$Ra})$= 1/2$, $\left<\sigma_N\right>$ is on the order of $\left\{0.1,\ldots,1\right\}$.
The resulting bound for the ${\cal{(CP)}}$-odd parameter is then
\begin{equation}
 |C_T| < 10^{-3}
\end{equation}
This limit is still about seven orders of magnitude weaker than the corresponding limit from measurements
\cite{Heckel_Hg_PRL2016} and calculations (see Table \ref{TAB:RT_HG}) on atomic Hg. However, the envisaged
experimental improvements on the Ra EDM measurement laid out in Ref. \cite{bishof_Ra_PRC2016} hold the 
promise to close this gap in future work.

\bibliographystyle{unsrt}
\newcommand{\Aa}[0]{Aa}


\end{document}